\documentclass[aps,prl,floatfix]{revtex4}
\usepackage{amssymb}
\usepackage{amsmath}
\usepackage{graphicx}

\begin{document}

\title{Magnetization of topological line-node semimetals}

\author{G.~P.~Mikitik}
\affiliation{B.~Verkin Institute for Low Temperature Physics \&
Engineering, Ukrainian Academy of Sciences, Kharkov 61103,
Ukraine}

\author{Yu.~V.~Sharlai}
\affiliation{B.~Verkin Institute for Low Temperature Physics \&
Engineering, Ukrainian Academy of Sciences, Kharkov 61103,
Ukraine}

\begin{abstract} Using an approximate expression for the Landau levels of the electrons located near a nodal line of a topological line-node semimetal, we obtain formulas for the magnetization of this semimetal at an arbitrary shape of its line.
It is also shown that the dependence of the chemical potential on the magnetic field can be strong in these materials, and this dependence  can essentially influence the de Haas - van Alphen oscillations. The obtained results are applied to the rhombohedral graphite which is one of the line-node semimetals. For this material, we find temperature and magnetic field dependences of its magnetic susceptibility.
\end{abstract}


\maketitle

\section{Introduction}

In recent years much attention has been given to the so-called topological semimetals \cite{wan,balents,young,sol,m-sh16,wang,neupane,borisenko,jeon,liang,ali,liu,z.wang,liu1,h.weng,
huang,S-Xu,Lv,shek,hei1,pie,weng,mullen,xie,kim,yu,yama,phil,schoop,neupane1,schoop1,huang1,bian,chen}.
In particular, it was predicted \cite{m-sh16} that their magnetization  exhibits unusual dependences on the chemical potential $\zeta$, the temperature $T$, and the magnetic field $H$.
These dependences  can serve as a fingerprint of the topological semimetals, and experimental investigations of the dependences can be useful in studying electron energy spectra of these materials.

There are several types of the topological semimetals.
In the Weyl and  Dirac semimetals, the electron energy bands contact at discrete points of the Brillouin zone and disperse linearly in all directions around these critical points. At present a number of such semimetals were discovered \cite{wang,neupane,borisenko,jeon,liang,ali,liu,z.wang,liu1,h.weng,
huang,S-Xu,Lv,shek}. The magnetization of these materials was theoretically analyzed in the papers \cite{m-sv,m-sh,kosh,roy,m-sh16}.
One more type topological materials is the line-node semimetals in which the conduction and valence bands touch along lines in the Brillouin zone and disperse linearly in directions perpendicular to these lines. It is necessary to emphasize that the contact of the electron energy bands along the lines is the widespread phenomenon in crystals \cite{herring,m-sh14,kim,fang}. For example, such contacts of the bands occur in graphite \cite{graphite}, beryllium \cite{beryl}, aluminium \cite{al}, and LaRhIn$_5$ \cite{prl04}. However, the degeneracy energy of the bands, $\varepsilon_d$, generally is not constant along such lines, and the $\varepsilon_d$ varies between its minimum $\varepsilon_{min}$ and maximum $\varepsilon_{max}$ values, reaching them at certain points of the line. A crystal with the band-contact line can be named the topological semimetal if the difference $\varepsilon_{max}- \varepsilon_{min}\equiv 2\Delta$ is sufficiently small
and if the chemical potential $\zeta$ of the electrons does not lie far away from the mean energy $\varepsilon_d^0 \equiv (\varepsilon_{max}+ \varepsilon_{min})/2$ of the line. Various line-node semimetals were theoretically predicted and discovered experimentally in recent years \cite{hei1,pie,weng,mullen,xie,kim,yu,yama,phil,schoop,neupane1,schoop1,huang1,bian,chen}.
The magnetic susceptibility of a crystal with a band-contact line characterized by large $\Delta$ was theoretically investigated many years ago \cite{m-sv,m-sh}. It turned out that the susceptibility exhibits a giant anomaly when $\zeta$ approaches one of the energies  $\varepsilon_{min}$ or $\varepsilon_{max}$ which correspond to the points of the electron topological transitions of $3\frac{1}{2}$ kind \cite{m-sh14}. When one deals with the topological semimetals, the interval $2\Delta$ is small, and the character of the anomaly in the susceptibility changes. The susceptibility in the case of the line-node semimetals was considered for weak magnetic fields in Ref. \cite{kosh15} and for arbitrary magnetic fields in Ref.~\cite{m-sh16}. However, in our paper \cite{m-sh16}, formulas for the magnetization were mainly obtained in the case of the semimetals with a closed band-contact line lying in a plane perpendicular to an axis of n-fold symmetry. Beside this, we did not consider the $H$-dependence of the chemical potential $\zeta$ and the effect of this dependence on the magnetization. But this dependence, as we shall see below, can be strong.

In this paper, we derive general formulas for the magnetization of a line-node semimetal with a band-contact line of an arbitrary shape, taking into account the dependence $\zeta(H)$. Then we apply these results to the case when the line terminates on  opposite faces of the Brillouin zone. As an example of the semimetal in which this situation occurs, we consider the rhombohedral graphite  \cite{hei1,pie,mcclure,kopnin}.

\section{Electron spectrum near a band-contact line}

\begin{figure}[tbp] 
 \centering  \vspace{+9 pt}
\includegraphics[scale=2]{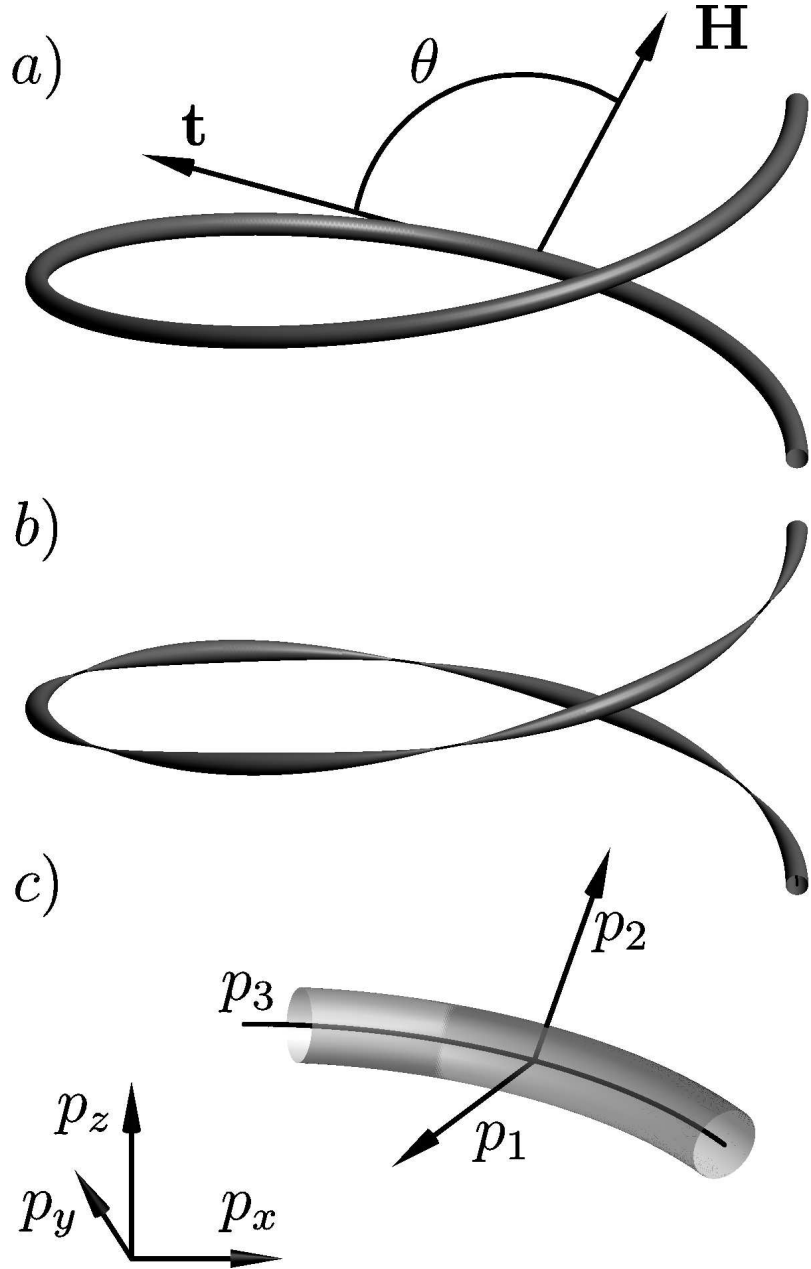}
\caption{\label{fig1} The Fermi surface of the rhombohedral graphite as an example of the Fermi surface in a  topological line-node semimetal at $|\zeta-\varepsilon_d^0|\gtrsim \Delta$ (a) and at $|\zeta-\varepsilon_d^0|< \Delta$ (b). The band-contact line lies inside the Fermi surface. The arrow shows the tangent vector ${\bf t}$ to the line at one of its points; $\theta$ is the angle between this ${\bf t}$ and the magnetic field ${\bf H}$. For clarity, we decrease the pitch of the helix. c) A part of the Fermi surface, of the band-contact line and the two coordinate systems connected with the band-contact line ($p_1$-$p_2$-$p_3$) and with crystallographic axes of the rhombohedral graphite ($p_x$-$p_y$-$p_z$).
 } \end{figure}

In the vicinity of a band-contact line along which the conduction and valance bands touch, let us introduce orthogonal curvilinear coordinates so that the axis ``$3$'' coincides with the line, Fig.~\ref{fig1}. The axes ``$1$'' and ``$2$'' are perpendicular to the third axis at every point of the band-contact line, and the appropriate coordinate $p_1$ and $p_2$ are measured from this line. In these coordinates, near the line, the most general form of the electron spectrum for the conduction and valence bands looks like \cite{m-sh14},
\begin{eqnarray}\label{1}
 \varepsilon_{c,v}\!\!&=&\!\varepsilon_d(p_3)\!+\!{\bf a}_{\perp}{\bf p}_{\perp}\pm E_{c,v},\\
 E_{c,v}^2\!\!&=&\!b_{11}p_1^2+b_{22}p_2^2, \nonumber
 \end{eqnarray}
where $\varepsilon_d(p_3)$ describes a dependence of the  degeneracy energy along the line (the $\varepsilon_{max}$ and $\varepsilon_{min}$ mentioned above are the maximum and minimum values of the function  $\varepsilon_d(p_3)$); ${\bf p}_{\perp}=(p_1,p_2,0)$ and ${\bf a}_{\perp}=(a_1,a_2,0)$ are the vectors perpendicular to the line; the parameters of the spectrum  $b_{11}$, $b_{22}$, and ${\bf a}_{\perp}$ generally depend on $p_3$. It is implied here that the directions of the axes ``1'' and ``2'' are chosen so that the quadratic form $E_{c,v}^2$ is diagonal (these directions generally changes along the line). The vector ${\bf a}_{\perp}$ specifies the tilt of the Dirac spectrum in the $p_1$-$p_2$ plane. Below we shall consider only the case when the length of the vector $\tilde{\bf a}_{\perp}\equiv (a_1/\sqrt{b_{11}},a_2/\sqrt{b_{22}},0)$ is less than unity \cite{comm1},
\begin{eqnarray*}
 \tilde a_{\perp}^2=\frac{a_1^2}{b_{11}}+\frac{a_2^2}{b_{22}}<1,
 \end{eqnarray*}
since at $\tilde a_{\perp}^2>1$ the magnetic susceptibility does not exhibit any essential anomaly in its dependences on $\zeta$, $H$, and $T$ \cite{m-sv,m-sh,m-sh16}.

When the parameter $\Delta \equiv (\varepsilon_{max} -\varepsilon_{min})/2$ is small as compared to the characteristic scale of electron band structure (i.e., $\Delta\ll 1$ eV) and $\tilde a_{\perp}^2<1$, the Fermi surface $\varepsilon_{c,v}({\bf p}_{\perp},p_3)=\zeta$ of the semimetal looks like a narrow electron or hole tube for $\zeta-\varepsilon_d^0\gtrsim \Delta$ or $\zeta- \varepsilon_d^0 \lesssim -\Delta$, respectively, Fig.~\ref{fig1}. The band-contact line lies inside this tube. If  $|\zeta- \varepsilon_d^0| <  \Delta$, the Fermi surface has a self-intersecting shape and consists of the electron and hole pockets touching at some points of the line, i.e., it looks like ``link sausages'', Fig.~\ref{fig1}. Thus, if the chemical potential $\zeta$ decreases and passes through the critical energies $\varepsilon_{max}=\varepsilon_d^0 + \Delta$ and $\varepsilon_{min}=\varepsilon_d^0 - \Delta$, the electron topological transitions occur \cite{m-sh14}. At these transitions, the electron tube first transforms into the self-intersecting Fermi surface and then this surface transforms into the hole tube. We shall assume below that all transverse dimensions of the Fermi-surface tubes and pockets, which are of the order of  $|\zeta- \varepsilon_d(p_3)|/V$  where $V\sim ({b_{11}b_{22}})^{1/4}$, are essentially less than the characteristic radius of curvature for the band-contact line. In this case  practically all electron orbits in the Brillouin zone, which are intersections  of the Fermi surface with planes perpendicular to the magnetic field, are small and lie near the band-contact line. In other words, a small region in the Brillouin zone determines the local  electron energy spectrum in the magnetic field almost for any point of the line. This spectrum has the form (see Appendix):
 \begin{eqnarray}\label{2}
\varepsilon_{c,v}^l(p_3)&=&\varepsilon_{d}(p_3) \pm \!\left(\frac{e\hbar\alpha H|\cos\theta|}{c}l\right)^{1/2}\!, \\
\alpha&=&\alpha(p_3)=2(b_{11}b_{22})^{1/2}(1-\tilde a_{\perp}^2)^{3/2}, \label{3}
 \end{eqnarray}
where $l$ is a non-negative integer ($l=0$, $1$, \dots),
with the single Landau subband $l=0$ being shared between the branches ``$c$'' and ``$v$'', and
$\theta=\theta(p_3)$ is the angle between the direction of the magnetic field and the tangent ${\bf t}={\bf t}(p_3)$ to the band-contact line at the point with the coordinate $p_3$, Fig.~\ref{fig1}. Formula (\ref{2}) fails only for those points of the line for which $\theta$ is close to $\pi/2$. However, these points do not give a noticeable contribution to the magnetization \cite{m-sv}.

\section{Magnetization}

We define the vector of the magnetization as $-\partial \Omega/\partial H_i$ where $\Omega$ is the $\Omega$ potential per unit volume of a crystal, and we disregard a contribution to the magnetization associated with the electron {\it surface} states of a topological semimetal. When the chemical potential $\zeta$ does not lie far away from $\varepsilon_d^0$, the total magnetization consists of its special part $M_i$ determined by the electron states located near the band-contact line and a background term $\chi_{ij}^0H_j$ in which the practically constant tensor $\chi_{ij}^0$ is specified by electron states located far away from this line,
\[
M_i^{\rm total}=M_i +\chi_{ij}^0 H_j.
\]
It is the special part $M_i$ that is responsible for dependences of the magnetization on the chemical potential, temperature, and for a nonlinear dependence of the magnetization on the  magnetic field magnitude. It is also significant that $|M_i|$ is not small as compared to $|\chi_{ij}^0 H_j|$ and can essentially exceed this background term for the topological semimetals \cite{m-sh16}. Below we calculate $M_i$ only.

In weak magnetic fields $H\ll H_T$, when the characteristic spacing  $\Delta\varepsilon_H$ between the Landau subbands is much less than the temperature $T$, the magnetization $M_i$ is proportional to $H$. On the other hand, at $H>H_T$, when $\Delta\varepsilon_H>T$, the magnetization becomes a nonlinear function of $H$. The background term $\chi_{ij}^0$ in the susceptibility remains constant at all magnetic fields. According to Eq.~(\ref{2}), we have the following estimate for the spacing $\Delta\varepsilon_H$ between the Landau subbands of electrons in the magnetic field: $\Delta\varepsilon_H\sim (e\hbar HV^2/c)^{1/2}$, and hence
\begin{equation*}
H_T\sim \frac{cT^2}{e\hbar V^2}.
\end{equation*}
If the characteristic velocity $V\approx ({b_{11}b_{22}})^{1/4}\sim 10^6-10^5$ m/s, one obtains $H_T\sim 2-200$ Oe  at $T=4$ K \cite{comm}. In other words, for the topological semimetals investigated at low temperatures, a nonlinear dependence of $M_i^{\rm total}$ on $H$ can develop at sufficiently low magnetic fields.

Using  formulas (\ref{2}) and (\ref{3}), the special part of the magnetization associated with the band-contact line can be calculated at magnetic fields of an arbitrary strength. In such  calculations, we shall suppose that Eqs.~(\ref{2}) and (\ref{3}) are valid at all angles $\theta$ including $\theta=\pi/2$. As was mentioned above [see also Eq.~(\ref{15})], this supposition does not introduce essential errors into the results. As is clear from formula (\ref{2}), within this approximation, a contribution of electron states located near a point $p_3$ to the special part of the $\Omega$ potential is determined only by the magnetic field component $H\cos[\theta(p_3)]$ parallel to the appropriate tangent ${\bf t}(p_3)$ to the line, and hence the magnetization of these states is parallel to this tangent, too. Eventually, we obtain the following expressions for the $\Omega$ potential and the magnetization at $T=0$ \cite{SuppMat}:
\begin{eqnarray}\label{4}
\Omega(\zeta,H)\!=\!-\frac{e^{3/2}H^{3/2}} {2\pi^2\hbar^{3/2}c^{3/2}}\!\!\int_{0}^{L}\!\!\!\!\!dp_3 |\cos\theta|^{3/2}\sqrt{\alpha(p_3)}K_1(u), \\
  {\bf M}(\zeta,H)\!=\!\frac{e^{3/2}H^{1/2}} {2\pi^2\hbar^{3/2}c^{3/2}}\!\!\int_{0}^{L}\!\!\!\!\!dp_3 |\cos\theta|^{1/2}\nu\!\sqrt{\alpha(p_3)} K(u){\bf t},\label{5}
  \end{eqnarray}
where the integration in the Brillouin zone is carried out over the band-contact line of the length $L$; $\theta=\theta(p_3)$ is angle between ${\bf t}={\bf t}(p_3)$ and ${\bf H}$; $\nu=\nu(p_3)$ is a sign of $\cos\theta$;
 \begin{eqnarray}\label{6}
 K_1(u)\!\!\!&=&\!\!\! \zeta(-\frac{1}{2},\![u]\!+\!1\!)+\sqrt{u}([u]+\frac{1}{2})- \frac{1}{3}u^{3/2}, \\ K(u)\!\!\!&=&\!\!\! \frac{3}{2}\zeta(-\frac{1}{2},\![u]\!+\!1\!)+\sqrt{u}([u]+\frac{1}{2}), \label{7}
  \end{eqnarray}
$\zeta(s,a)$ is the Hurwitz zeta function,
 \begin{equation}\label{8}
 u=\frac{[\zeta-\varepsilon_d(p_3)]^2 c}{e\hbar  \alpha(p_3) H|\cos\theta|}=\frac{cS(p_3)}{2\pi e\hbar H},
 \end{equation}
$S(p_3)$ is the area of the Fermi-surface cross section
by the plane perpendicular to the magnetic field and passing through the point with the coordinate $p_3$, $[u]$ is the integer part of $u$ ($[u]$ is the number of the Landau levels lying below $\zeta$ at the point $p_3$). In deriving Eqs. (\ref{4}) and (\ref{5}), we have assumed the two-fold degeneracy of the electron bands in spin. In absence of this degeneracy (for a noncentrosymmetric semimetal with a strong spin-orbit interaction), the right hand sides of formulas (\ref{4}) and (\ref{5}) should be divided by two. In the case of a closed band-contact line, formula (\ref{5}) reproduces  Eqs.~(44), (46), (47) of Ref. \cite{m-sh16}. For nonzero $T$, the $\Omega$ potential and the magnetization $M_i(\zeta,H,T)$ can be calculated with the relationships \cite{rum}:
 \begin{eqnarray}\label{9}
 \Omega(\zeta,H,T)=-\int_{\!-\infty}^{\infty}\!\! d\varepsilon \Omega(\varepsilon,H,0)f'(\varepsilon), \\
 M_{i}(\zeta,H,T)=-\int_{\!-\infty}^{\infty}\!\! d\varepsilon M_{i}(\varepsilon,H,0)f'(\varepsilon),\label{10}
 \end{eqnarray}
where $f'(\varepsilon)$ is the derivative of the Fermi function,
\begin{equation}\label{11}
 f'(\varepsilon)=-\left[4T\cosh^2\left(\frac{\varepsilon- \zeta}{2T}\right)\right]^{-1}.
 \end{equation}

In the topological semimetals, charge carriers (electron and holes) are located near the band-contact line, and their chemical potential $\zeta$ generally depends on the magnetic field, $\zeta=\zeta(H)$. This dependence can be derived from the condition that the charge carrier density $n$ does not vary with increasing $H$,
\begin{equation}\label{12}
n(\zeta,H)=n_0(\zeta_0),
\end{equation}
where $n_0$ and $\zeta_0$ are the density and the chemical potential at $H=0$,
 \[
 n(\zeta,H)=n_0(\zeta)-\frac{\partial \Omega}{\partial \zeta},
 \]
and $\Omega$ is given by Eq.~(\ref{4}). With Eqs.~(\ref{1}), (\ref{2}) and (\ref{4}), one finds  the following expression for $n_0(\zeta_0)$ and $n(\zeta,H)$ at $T=0$:
\begin{eqnarray}\label{13}
 n_0(\zeta_0)=\frac{1}{2\pi^2\hbar^3}\int_{0}^{L}\!\!\!\!\!dp_3
\frac{(\zeta_0-\varepsilon_d(p_3))^2\sigma(\zeta_0- \varepsilon_d(p_3))}{\alpha(p_3)}, \\
 n(\zeta)=\frac{eH}{2\pi^2 c\hbar^2}\int_{0}^{L}\!\!\!\!\!dp_3
|\cos\theta|\sigma(\zeta-\varepsilon_d(p_3)) (\frac{1}{2}+[u]),
\label{14}
 \end{eqnarray}
where $\sigma(x)=1$ if $x>0$, and $-1$ otherwise. The other notations are the same as in formulas (\ref{4})--(\ref{8}). At nonzero temperatures, $n_0(\zeta_0,T)$ and $n(\zeta,H,T)$ can be calculated  with formulas similar to Eq.~(\ref{9}). On calculating $\zeta(H)$ with Eqs.~(\ref{12})--(\ref{14}), one can find the magnetization as a function of $n_0$ or $\zeta_0$, inserting $\zeta(H)$ into Eq.~(\ref{5}).

Consider now several limiting cases.
In the weak magnetic field, $H\ll H_T$, the $\Omega$ potential described by formulas (\ref{4}) and (\ref{9}) becomes proportional to $H^2$ \cite{SuppMat}. Eventually, we arrive at linear dependence of the magnetization on the magnetic field,
\begin{eqnarray}\label{15}
  {\bf M}(\zeta,H,T)\!=\!\frac{e^{2}H} {12\pi^2\hbar c^{2}}\!\!\int_{0}^{L}\!\!\!\!\!dp_3 \alpha(p_3)f'(\varepsilon_d)\cos\theta\,{\bf t}.
  \end{eqnarray}
This formula agrees with Eq.~(35) of Ref. \cite{m-sh16}. Note also  that points of the band-contact line for which $\theta$ is close to $\pi/2$ give a small contribution to the magnetization.
Equation (\ref{15}) leads to the following expression for the magnetization component $M_{\parallel}$ parallel to the magnetic field:
\begin{eqnarray*}
   M_{\parallel}(\zeta,H,T)\!=\!\frac{e^{2}H} {12\pi^2\hbar c^{2}}\!\!\int_{0}^{L}\!\!\!\!\!dp_3 \alpha(p_3)f'(\varepsilon_d)\cos^2\theta.
  \end{eqnarray*}
Interestingly, this expression can be easily understood from the following considerations: At a given $p_3$, the Landau levels  described by Eq.~(\ref{2}) look like the levels of electrons near the Dirac point of graphene, $\varepsilon_l=\pm (2e\hbar V_D^2Hl/c)^{1/2}$, if the energy of the Dirac point coincides with $\varepsilon_d(p_3)$, and the electron velocity $V_D$ at this point is given by $V_D^2=\alpha |\cos\theta|/2$. In the weak magnetic field $H$ perpendicular to the ``graphene'' plane, the magnetic moment of the electrons near the Dirac point point has the form \cite{kosh07}:
\begin{eqnarray*}
   \frac{e^{2}V_D^2 H} {3\pi c^{2}}f'(\varepsilon_d)=\frac{e^{2}\alpha|\cos\theta| H} {6\pi c^{2}}f'(\varepsilon_d).
    \end{eqnarray*}
Multiplying this expression by $dp_3 |\cos\theta|/2\pi\hbar$ (the number of the ``graphene'' planes in the interval $dp_3$) and  integrating over the band-contact line, we arrive at the above formula for $M_{\parallel}$.

In strong magnetic fields, $T\ll \Delta\varepsilon_H$, and if  $|\zeta-\varepsilon_{min}|$, $|\zeta-\varepsilon_{max}|\ll \Delta\varepsilon_H$, the argument $u$ in $K(u)$ is small practically for all points of the band-contact line, and hence $K(u)\approx (3/2)\zeta(-1/2,1)\approx -0.98/\pi$. In this case, formula (\ref{5}) gives $M_i\propto H^{1/2}$ with a proportionality coefficient depending on the direction of the magnetic field and the shape of the band-contact line.
In particular, we find the following expression for longitudinal component of the magnetization:
\begin{eqnarray*}
   M_{\parallel}(H)=\frac{3\zeta(-1/2,1)e^{3/2}H^{1/2}} {4\pi^2\hbar^{3/2}c^{3/2}}\!\!\int_{0}^{L}\!\!\!\!\!dp_3 |\cos\theta|^{3/2}\sqrt{\alpha(p_3)}.
  \end{eqnarray*}
As in the case of weak magnetic fields, this formula can be  represented in the form: $\int_0^L dp_3 |\cos\theta|M_D/2\pi\hbar$,  where $M_D$ is the electron magnetic moment of the Dirac point in strong magnetic fields \cite{sharapov},
\begin{eqnarray*}
   M_D\!\!=\!\frac{3\zeta(\!-\!1\!/\!2,\!1)e^{3\!/2}V_D\! H^{\!1\!/2}} {\sqrt{2}\pi\hbar^{1/2}c^{3/2}}\!=\! \frac{3\zeta(\!-\!1\!/\!2,\!1)e^{3\!/2}\!(\alpha|\!\cos\theta|H\!)^{1\!/2}} {2\pi\hbar^{1/2}c^{3/2}}.
  \end{eqnarray*}
As to condition (\ref{12}), in strong magnetic field and at low temperatures, it leads  to a shift of the chemical potential into the interval: $\varepsilon_{min} < \zeta < \varepsilon_{max}$.

In the region of the magnetic fields when $T\ll \Delta\varepsilon_H \ll |\zeta-\varepsilon_{min}|$, $|\zeta-\varepsilon_{max}|$, $2\Delta$, it follows from Eq.~(\ref{5}) \cite{SuppMat} that the magnetization is described by the usual formula \cite{Sh} for the de Haas - van Alphen effect, with the phase of the oscillations being shifted by $\pi$ \cite{shen}. This shift is the characteristic feature of crystals with a band-contact line and is due to the Berry phase $\pi$ for the electron orbits surrounding this line \cite{prl,shen}. In the equivalent interpretation \cite{jetp,g1} allowing a nonzero spin-orbit coupling, this shift is caused by the large value of the orbital $g$ factor, $g=2m/m_*$, occurring even at a weak spin-orbit interaction. Here $m_*$ is the cyclotron mass and $m$ is the electron mass. As in usual metals \cite{Sh}, the dependence $\zeta(H)$ is sufficiently weak in this region of the magnetic fields and practically has no effect on the oscillations.

\begin{figure}[tbp] 
 \centering  \vspace{+9 pt}
\includegraphics[scale=2]{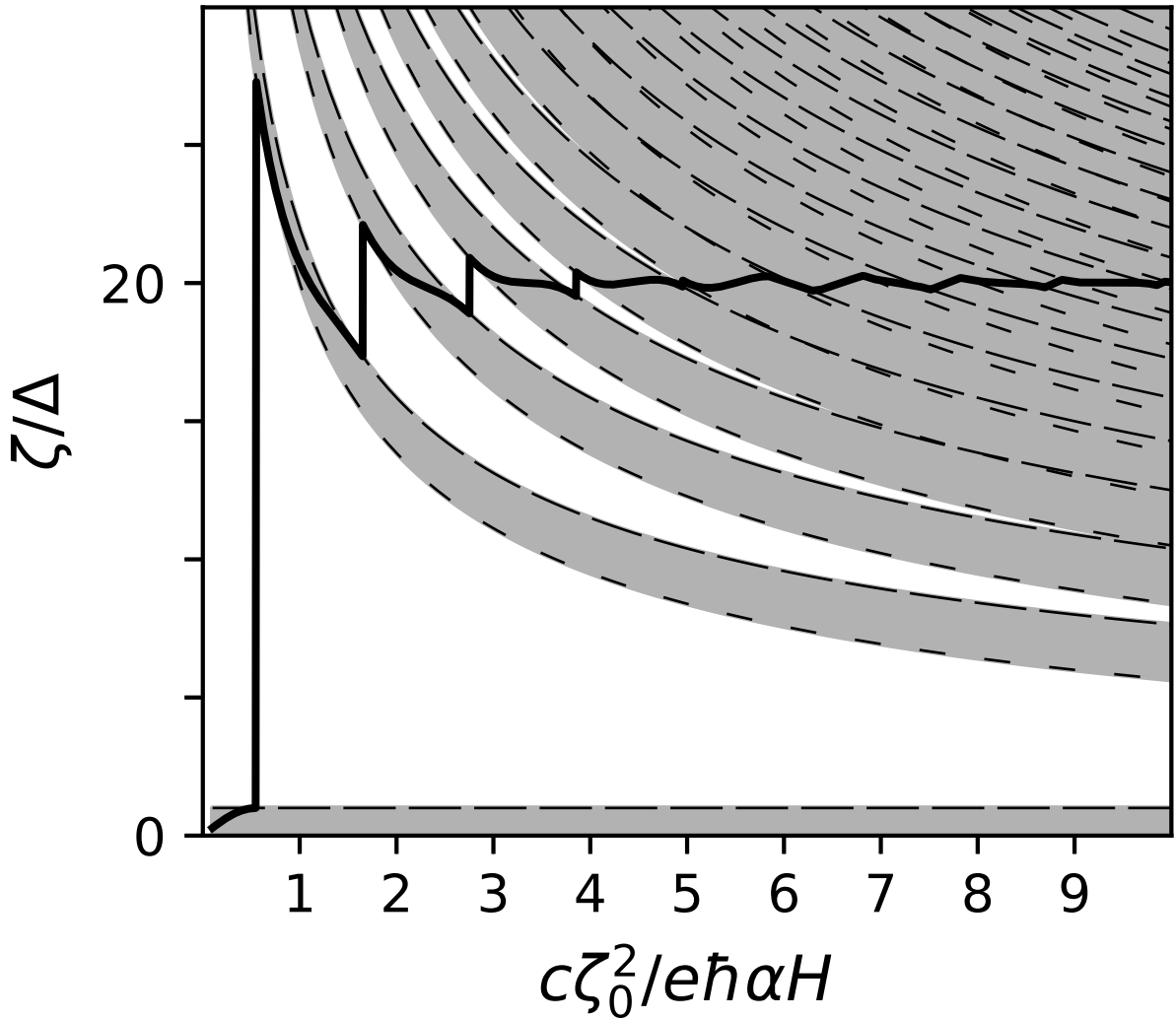}
\caption{\label{fig1a} The dependence of chemical potential $\zeta$ (measured from $\varepsilon_d^0$) on $1/H$ calculated with Eqs.~(\ref{12})--(\ref{14}) at $\zeta(H=0)\equiv \zeta_0=20\Delta$, $\varepsilon_d(p_3)=\Delta \cos(2\pi p_3/L)$, $\cos\theta=1$, $\alpha(p_3)=$const. We also mark the Landau subbands, Eq.~(\ref{2}), by the dark background, and the short and long dashes indicate the lower and the upper edges of these subbands, respectively. The crossover described in the text occurs at $c\zeta_0^2/e\hbar\alpha H\sim 5$.
 } \end{figure}   

At small $\Delta$ when
$T< 2\Delta \ll \Delta\varepsilon_H \ll |\zeta-\varepsilon_{min}|$, $|\zeta-\varepsilon_{max}|$, the spectrum (\ref{2}) transforms, in fact, into the spectrum of a two-dimensional electron system since different Landau subbands $\varepsilon_{c,v}^l(p_3)$ do not overlap, and they look like broadened Landau levels. In this case, when $H$ changes, the chemical potential $\zeta(H)$ moves together with one of these levels, and then, at a certain value of $H$, it jumps from this level to the neighboring one \cite{Sh}, Fig.~\ref{fig1a}. This strong dependence $\zeta(H)$ noticeably changes the shape of the de Haas - van Alphen oscillations (see the next section) and can mask the correct value of the Berry phase when it is measured with these oscillations. Indeed, the jumps occur at the fields $H_l$ for which $n(\zeta)$ in Eq.~(\ref{14}) becomes independent of $\zeta$. This situation is realized when $[u]$ in the right hand side of Eq.~(\ref{14}) is one and the same integer $l$ along the whole line. Then, Eq.~(\ref{12}) takes the form:
\begin{equation*}
\frac{1}{H_l}=\frac{eC}{2\pi^2c\hbar^2}\left (l+\frac{1}{2}\right ),
\end{equation*}
where the constant $C$ is the ratio of $\int_0^Ldp_3|\cos\theta|$ to $n_0(\zeta_0)$. It follows from this equation that the dependence of $1/H_l$ on $l$ is a straight line that intersects the $l$ axis at $l=-1/2$, i.e, the Landau-level fan diagram plotted with the fields $H_l$ looks like in the case when the Berry phase $\Phi_B$ is equal to zero \cite{Sh,shen}. If $\Delta\varepsilon_H$ decreases and becomes comparable with $2\Delta$, a crossover from the quasi-two-dimensional electron spectrum to the three-dimensional one occurs, the jumps in $\zeta$ smooth, and the appropriate Landau-level fan diagram can give an intermediate value of the ``Berry phase'' lying between $0$ and $\pi$.

Strictly speaking, the quasi-two-dimensional electron spectrum in magnetic fields does not appear in every topological semimetal with a small $\Delta$ since if $\cos\theta\to 0$ in  some part of the line, the $\Delta\varepsilon_H$ becomes less than $2\Delta$ there; see Eq.~(\ref{2}). For the quasi-two-dimensional spectrum to occur, a change of the quantity $u\propto 1/\cos\theta$ along the line should be less than unity, i.e., $1/\cos\theta$ may change only within a sufficiently small interval. This imposes a restriction on the shape of the nodal line. It is clear that the spectrum of this kind can appear for a straight band-contact line, i.e., for a symmetry axis, since $\theta(p_3)$ is constant in this case. As will be shown in the next section, the quasi-two-dimensional spectrum is also possible in the case of  band-contact lines terminating on the opposite faces of the Brillouin zone for a certain region of the magnetic-field directions. This type of the spectrum can also occur for a closed band-contact line composed of nearly straight arcs. This situation appears to take place in ZrSiS \cite{schoop,neupane1}. Besides, the spectra including the quasi-two dimensional and three-dimensional parts can appear in the line-node semimetals containing several small groups of charge carriers. In this case one may expect to find a noticeable dependence of $\zeta$ on $H$ and to obtain the intermediate values of $\Phi_B$ in the measurements of the de Haas - van Alphen oscillations.

Using various oscillation effects observed in magnetic fields, the Berry phase was recently found in the experiments with ZrSiS \cite{Ali1,wang1,Hu1,pez,singha,matus} and with ZrSiTe or ZrSiSe \cite{Hu}, and the intermediate values of this phase (other than $0$ and $\pi$) were obtained for a number of the electron orbits.
Taking into account the above considerations, one may hypothesize  that the essential dependence of chemical potential on the magnetic field  takes place in these experiments. This dependence is probably associated with the existence of an electron group for which the quasi-two-dimensional spectrum or the crossover to this spectrum occurs in the magnetic-field range under study in these semimetals.

For comparison, let us discuss the well-known measurements of the Berry phase in graphene \cite{nov05,zhang05}. These measurements revealed the genuine Berry phase $\pi$ for the electron orbits surrounding the Dirac point, even though one might expect a strong dependence of the chemical potential on $H$ in this two-dimensional material. However, the oscillation experiments described in Refs.~\cite{nov05,zhang05} were carried out at fixed values of the gate voltage. This means that the measurements were made at constant chemical potential rather than at constant number of the charge carriers, and so the true value of the Berry phase was found in these experiments.

Finally, it is worth noting that the obtained results for the magnetization can be useful in describing the magnetostriction of the topological semimetals \cite{fnt17}.

\section{Rhombohedral graphite}

We now apply the above results to the rhombohedral graphite \cite{hei1,pie,mcclure,kopnin}. According to Ref. \cite{kopnin}, in this material there is a band-contact line that has the shape of a helix terminating on the opposite faces of the Brillouin zone, Fig.~1. In the simplest model of Ref.~\cite{kopnin}, the helix is described as follows:
 \begin{eqnarray}
p_x=p_0\cos\phi,\ \ \ p_y=p_0\sin\phi, \nonumber \\
p_z=\frac{\hbar}{d}(\phi-\frac{\pi}{6}), \label{16}
 \end{eqnarray}
where the $p_x$-$p_y$ plane of the quasimomentum space coincides with the basal plane of the crystal, and the third component of the quasimomentum, $p_z$, is perpendicular to this plane; $\phi$ is the angle defining the direction of the quasimomentum in the $p_x$-$p_y$ plane; $d\approx 3.35$ {\AA} is the interlayer distance in the rhombohedral graphite, and $p_0=\gamma_1/v_F$ is a constant, with $\gamma_1\approx 0.39$ eV, $v_F=1.5\gamma_0a_0/\hbar \approx 1.04\cdot 10^6$ m/sec, $a_0=1.42$ {\AA}, and  $\gamma_0=3.2$ eV being the parameters of the model. Within this model, the electron spectrum near the helicoidal band-contact line reduces to Eq.~(\ref{1}) with $\varepsilon_d(p_3)=0$ (from here on, we measure electron energies from the energy of the band degeneracy), ${\bf a}_{\perp}=0$, and
\begin{eqnarray} \label{17}
 b_{11}&=&v_F^2, \nonumber \\
 b_{22}&=&v_F^2(1+\tilde p_0^2), \\
 \alpha(p_3)&=&2v_F^2 \sqrt{1+\tilde p_0^2}, \nonumber
 \end{eqnarray}
where $\tilde p_0 \equiv p_0 d/\hbar\approx 0.19$.
Thus, in the model, $b_{11}$, $b_{22}$ and $\alpha(p_3)$ are constant along the line, and the parameter $\Delta\equiv \varepsilon_{max}- \varepsilon_{min}=0$.
With formulas (\ref{16}), one can find the tangent vector ${\bf t}$ to the line,
 \begin{eqnarray}\label{18}
 {\bf t}=\frac{1}{\sqrt{1+\tilde p_0^2}}(-\tilde p_0\sin\phi, \tilde p_0\cos\phi, 1),
 \end{eqnarray}
and the infinitesimal element $dp_3$,
 \begin{eqnarray}\label{19}
 dp_3=\frac{\hbar \sqrt{1+\tilde p_0^2}}{d}d\phi,
 \end{eqnarray}
which are both expressed in terms of the angle $\phi$.

Let the magnetic field ${\bf H}$ have the components:
\[
{\bf H}=H(\sin\theta_H \cos\phi_H,\sin\theta_H \sin\phi_H, \cos\theta_H),
\]
where the angles $\theta_H$ and $\phi_H$ define its direction relative to the crystal axes $x$, $y$, and $z$.
Then, a simple calculation gives
\begin{eqnarray}\label{20}
\cos\theta &=&\frac{1}{\sqrt{1+\tilde p_0^2}}\lambda(\theta_H,\varphi),  \\
u&=&\frac{\zeta^2 c}{2e\hbar v_F^2 |\lambda(\theta_H,\varphi)|H}, \label{21}
 \end{eqnarray}
where $\varphi=\phi-\phi_H$,
\begin{eqnarray}\label{22}
 \lambda(\theta_H,\varphi) =\cos\theta_H-\tilde p_0 \sin\theta_H \sin\varphi.~~~
  \end{eqnarray}

\begin{figure}[tbp] 
 \centering  \vspace{+9 pt}
\includegraphics[scale=2]{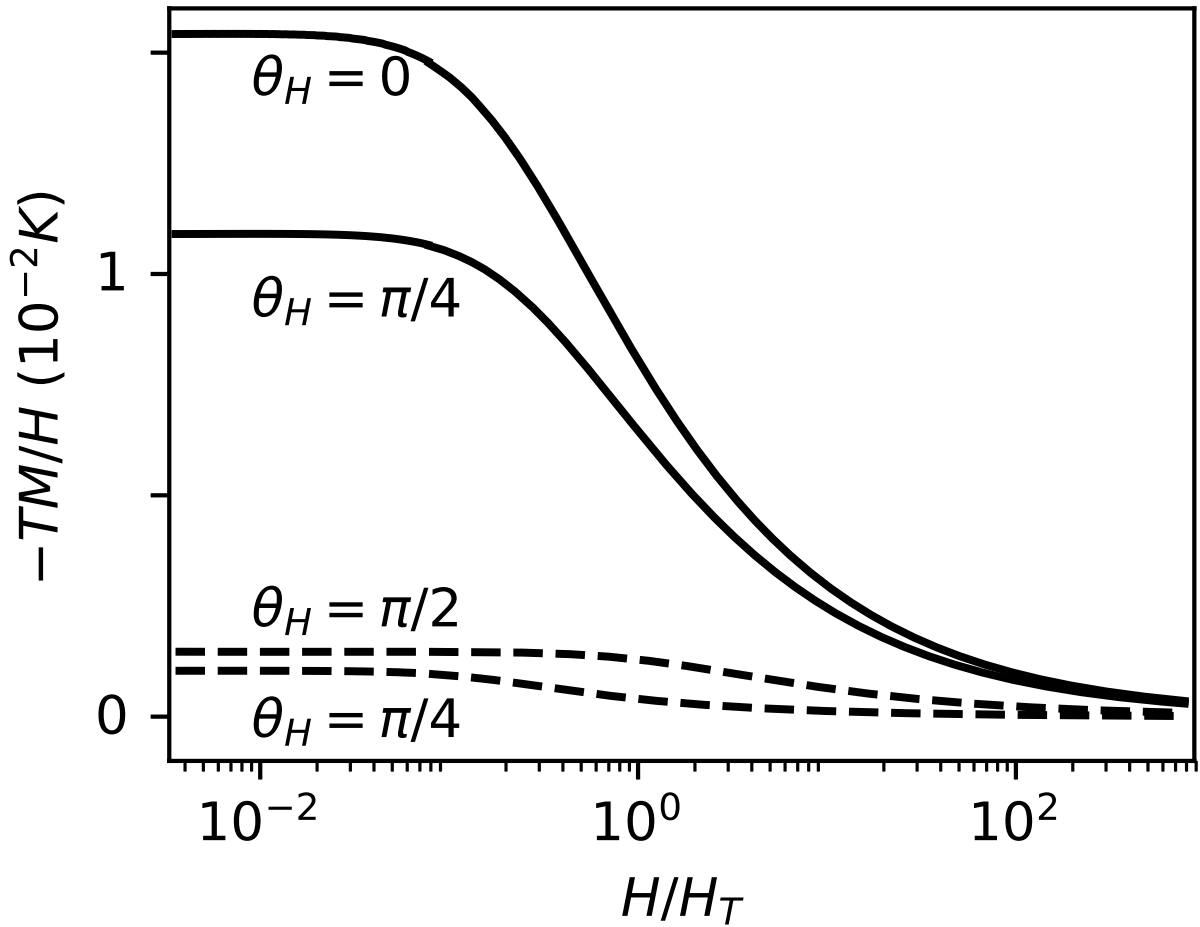}
\caption{\label{fig2} Dependences of $TM_z/H$ (solid lines) and of $TM_{xy}/(\tilde p_0 H)$ (dashed lines) on $H/H_T$ calculated numerically with Eqs.~(\ref{10}), (\ref{23}), and (\ref{24}) at $\zeta=0$ and $\theta_H=0$, $\pi/4$ for $M_z$ and  $\theta_H=\pi/2$, $\pi/4$ for $M_{xy}$. The $H_T$ is given by Eq.~(\ref{26}). At $H/H_T\gg 1$, the combinations $TM_z(H)/H$ and $TM_{xy}(H)/H$ are proportional to $(H_T/H)^{1/2}$.
 } \end{figure}   

With Eqs.~(\ref{17})--(\ref{22}), and (\ref{5}), we obtain the following expressions for the magnetization components at $T=0$:
\begin{eqnarray}\label{23}
 {\bf M}_{xy}\!\!\!\!&=&\!\!\!-\frac{\tilde p_0 v_F H^{1/2}}{\sqrt 2 \pi^2 d}\!\left(\frac{e}{\hbar c}\right)^{\!3/2}\!\!\!{\bf h}_{\perp}\!\!\!\int_{-\pi}^{\pi}\!\!\!\!d\varphi \sin\varphi |\lambda|^{1/2}\!\nu_{\lambda}K(u),~~~~  \\
M_z&=&\frac{v_F H^{1/2}}{\sqrt 2 \pi^2 d}\left(\frac{e}{\hbar c}\right)^{\!3/2}\!\!\!\int_{-\pi}^{\pi}\!\!\!d\varphi |\lambda|^{1/2}\!\nu_{\lambda}K(u), \label{24}
  \end{eqnarray}
where $\lambda=\lambda(\theta_H,\varphi)$, $\nu_{\lambda}$ is the sign of $\lambda$, the direction of the component $M_{xy}$ in the $x$-$y$ plane is determined by the unit vector ${\bf h}_{\perp}=(\cos\phi_H,\sin\phi_H,0)$, $u$ is given by Eq.~(\ref{21}). With Eqs.~(\ref{23}) and (\ref{24}), one can also calculate the magnetic torque $K=H(M_z\sin\theta_H - M_{xy}\cos\theta_H)$.
For weak magnetic fields, Eq.~(\ref{15}) yields
\begin{eqnarray}\label{25}
 {\bf M}\!\!=\!\!\frac{e^2v_F^2 H}{6\pi c^2 d}f'(0)\!\left(\tilde p_0^2 \sin\theta_H {\bf h}_{\perp} +2\cos\theta_H {\bf z}\right),
  \end{eqnarray}
where ${\bf z}$ is the unit vector along $z$ axis. Since $\tilde p_0^2\approx 0.04$, this formula shows that at $\frac{\pi}{2}-\theta_H \gg 0.02$, the magnetization is mainly directed along $z$ axis, and $|M_z/H_z|$ can reach a large value of the order of $0.016/T$ where $T$ is measured in Kelvin.

The background susceptibility tensor $\chi_{ij}^0$ for rhombohedral  graphite has the two components: $\chi_{zz}^0$ and $\chi_{xx}^0=\chi_{yy}^0=\chi_{\perp}^0$, and hence the total magnetization ${\bf M}^{\rm total}$ is described by the formulas:
\begin{eqnarray*}
M_z^{\rm total}&=&M_z+\chi_{zz}^0 H\cos\theta_H,\\
M_{xy}^{\rm total}&=&{\bf M}_{xy}+\chi_{\perp}^0 H\sin\theta_H {\bf h}_{\perp}.
\end{eqnarray*}
The constants $\chi_{zz}^0$ and $\chi_{\perp}^0$ are independent of  the temperature and the magnetic field, and so the background terms have no effect on dependences of the total magnetization on $H$ and $T$. But it is well to bear in mind that
these terms can be  generally essential in analyzing experimental $\theta_H$-dependences of $M_z^{\rm total}$ and $M_{xy}^{\rm total}$.

\begin{figure}[tbp] 
 \centering  \vspace{+9 pt}
\includegraphics[scale=2]{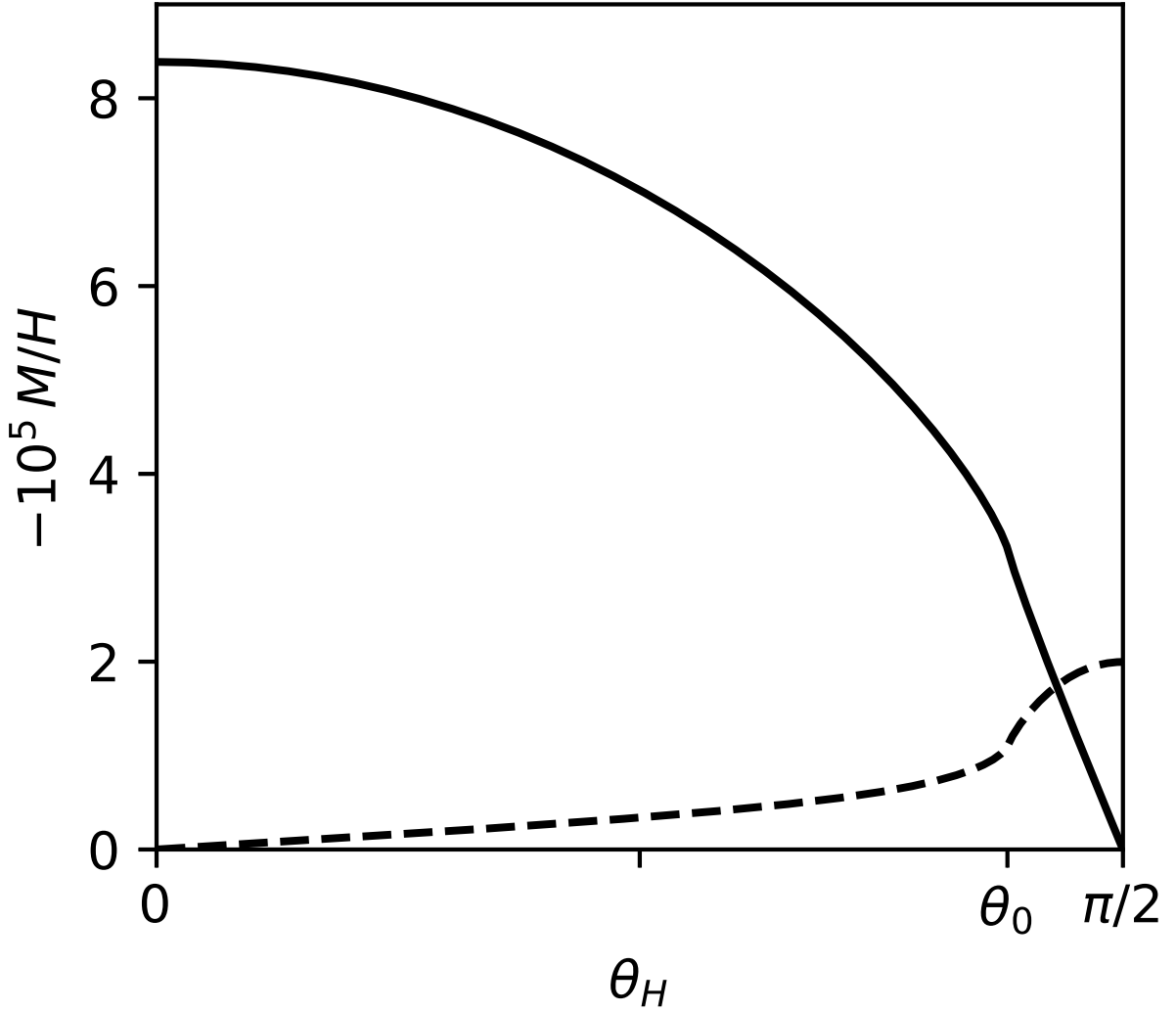}
\caption{\label{fig3} Dependences of $M_z/H$ (the solid line) and of $M_{xy}/(\tilde p_0H)$ (the dashed line) on the tilt angle $\theta_H$ of the magnetic field $H=10$ T. The dependences are calculated numerically with Eqs.~(\ref{10}), (\ref{23}), and (\ref{24}) at the temperature $4.2$ K and $\zeta=0$. The angle $\theta_0$ is defined by Eq.~(\ref{27}). With decreasing $H$, the dependences remain qualitatively unchanged, but values of $M_z(0)/H$ and $M_{xy}(\pi/2)/H$ increase in agreement with Fig.~\ref{fig2}.
 } \end{figure}

Consider the case of the ``ideal'' semimetal (without any doping) when $\zeta= 0$. In this situation, it follows from formulas (\ref{10}), (\ref{23}), and (\ref{24}) that for any component $M_i$ of the magnetization, the combination $T M_i/H$ depends only on the direction of the magnetic field and on the ratio $H/T^2$, i.e. on $H/H_T$ where we define $H_T$ from the condition $|\varepsilon_{c,v}^1-\varepsilon_{c,v}^0|_{\theta=0}=T$,
\begin{equation}\label{26}
H_T=\frac{cT^2}{2e\hbar v_F^2\sqrt{1+\tilde p_0^2}}.
\end{equation}
In Fig.~\ref{fig2} we show the $H$-dependences of the combinations  $TM_z/H$ and $TM_{xy}/(\tilde p_0 H)$ for $\theta_H=0$, $\pi/4$, and $\theta_H=\pi/4$, $\pi/2$, respectively. It is seen that at weak fields $H\ll H_T$, the magnitude of the magnetic  susceptibility $|M_i/H|$ is maximum and is proportional to $1/T$ in agreement with Eq.~(\ref{25}). For strong magnetic fields $H\gg H_T$, we find that $TM_i/H \propto (H_T/H)^{1/2}$, i.e., $M_i \propto H^{1/2}$ in accord with the result of the previous section. At fixes $H$ and $T$, the angular dependences of $M_z$ and $M_{xy}$ are shown in Fig.~\ref{fig3}. The component $M_z$ is zero  when the magnetic field lies in the $x$-$y$ plane, whereas $M_{xy}$ vanishes at $\theta_H=0$. The characteristic angle $\theta_0$ visible as a crossover point in the plot is determined by the equality,
 \begin{eqnarray}\label{27}
\cot\theta_0=\tilde p_0.
 \end{eqnarray}
The origin of this crossover is the following:
At $\theta_H>\theta_0$, points in the band-contact line exist for which the tangent to the line is perpendicular to the magnetic field, whereas for $\theta_H<\theta_0$, such points are absent. In our approximation these points do not give any contribution to the magnetization, and the appearance of these points at $\theta_H>\theta_0$ leads to the crossover.

\begin{figure}[tbp] 
 \centering  \vspace{+9 pt}
\includegraphics[scale=1.5]{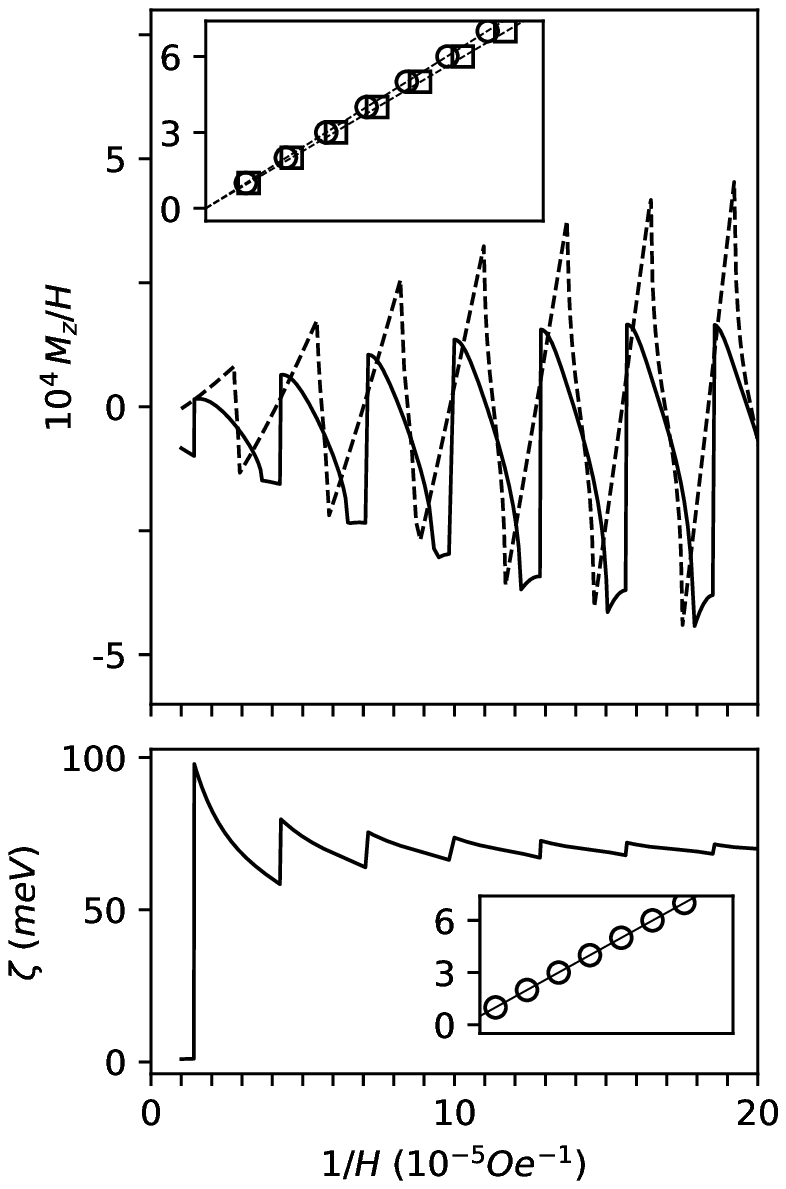}
\caption{\label{fig4} Top: Dependences of $M_z/H$ on $H$ at $T=0$ and the magnetic field directed along $z$-axis. The dependences are  calculated numerically with Eqs.~(\ref{5}), (\ref{7}), (\ref{8}), (\ref{17})--(\ref{20}), (\ref{22}) and (\ref{28}). The dashed line corresponds to the constant chemical potential $\zeta=70$ meV, the solid line shows $M_z/H$ at $\zeta_0=70$ meV, taking into account the $H$-dependence of $\zeta$ presented in the bottom panel. The inset: The Landau-level fan diagram plotted with the positions of the maxima (circles) and the minima (squares) of the dashed curve in the main top panel. Bottom: The $H$-dependence of the chemical potential calculated with Eqs.~(\ref{12})--(\ref{14}) at $\zeta_0=70$ meV. The inset shows the Landau-level fan diagram plotted with the positions of the maxima of the solid curve in the top panel.
} \end{figure}   

Of course, for real samples of the rhombohedral graphite, one cannot expect that the doping is completely absent, and $\zeta=0$. Besides, as was mentioned in the Introduction, the parameter $\Delta$ always differs from zero in the line-node semimetals. A more accurate model of Ref.~\cite{kopnin} for electron energy spectrum of the rhombohedral  graphite shows that $\varepsilon_d$ has the form:
 \begin{eqnarray}\label{28}
\varepsilon_d\approx -\Delta \cos(3p_zd/\hbar)=-\Delta\sin 3\phi,
 \end{eqnarray}
where $\Delta\approx 2\gamma_4 \gamma_3 \gamma_1/\gamma_0^2 \approx 1$ meV,  $\gamma_4=44$ meV, and $\gamma_3 \approx 315$ meV.
The data of Figs.~\ref{fig2} and \ref{fig3} will remain unchanged if $\zeta$ and $\Delta$ do not exceed $T$ or $\Delta\varepsilon_H$, i.e., if $\Delta, \zeta \ll {\rm max}(T,\Delta\varepsilon_H)$ where $\Delta\varepsilon_H [{\rm meV}]\sim 0.4\sqrt{H[{\rm Oe}]}$.
In Fig.~\ref{fig4} we present the $H$-dependence of $M_z/H$ when at least $\zeta$ does not satisfy this restriction. The de Haas - van Alphen oscillations are clearly  visible in the figure. The dashed line shows the oscillations calculated at constant $\zeta$. Since the phase of these oscillations is determined by the Berry phase $\Phi_B$ for the appropriate electron orbits \cite{shen}, this $\Phi_B$ can be found with the Landau-level fan diagram shown in  the upper inset of Fig.~\ref{fig4}. (As expected, this inset yields $\Phi_B=\pi$.) However, $\Delta\varepsilon_H$ at the Fermi level exceeds $2\Delta$
for $H> H_{cr}$ where $H_{cr}\sim 4\Delta\zeta c/e\hbar \alpha \sim 2$ kOe is the crossover field separating the regions of the quasi-two-dimensional and three-dimensional regimes of the oscillations. In other words, for the whole interval of the magnetic fields presented in Fig.~\ref{fig4}, the electron spectrum is quasi-two-dimensional, and one has $\delta u <1$ for such magnetic fields where $\delta u \approx (4\Delta/\zeta)u$ is the variation of $u$ along the band-contact line. Then, as  explained in the previous section, the chemical potential exhibits jumps when the middle of the interval between the appropriate Landau subbands crosses $\zeta_0$, the chemical potential at zero magnetic field, Fig.~\ref{fig4}. At these $H$, the magnetization $M_z$ experiences jumps, too, and the positions of these jumps do not coincide with the sharp peaks of $M_z/H$ calculated at a constant chemical potential. In other words, there is a shift of the oscillations associated with the dependence $\zeta(H)$, and this shift imitates a change of the Berry phase $\Phi_B$. In particular, the lower inset in Fig.~\ref{fig4} suggest that $\Phi_B=0$ although the Berry phase is still equal to $\pi$ for the electron orbits in the magnetic fields. Note also that when the dependence $\zeta(H)$ is taken into account, the shape of the oscillations essentially changes as compared to the shape of the oscillations calculated at $\zeta=$ const.

If the magnetic field is tilted away from the $z$ axis, the $\cos\theta$ is no longer constant along the line, and the crossover from the three-dimensional electron spectrum to the quasi-two-dimensional one develops at the higher magnetic fields  than $2$ kOe. For the quasi-two-dimensional spectrum to occur, it is necessary that $(\delta\cos\theta/\cos\theta)u < 1$ where $\delta\cos\theta$ is the variation of $\cos\theta$ along the line. Using Eqs.~(\ref{20})--(\ref{22}), we obtain $(\delta\cos\theta/\cos\theta)\approx 2\tilde p_0\tan\theta_H$ and apart from $(4\Delta/\zeta)u<1$, the additional condition on $H$: $c\zeta^2\tilde p_0\tan\theta_H/(e\hbar  V_F^2\cos\theta_H) < H$. At $\zeta=70$ meV, this condition has no effect on the crossover field $H_{cr}\sim 2$ kOe when $\theta_H \lesssim 10^{\circ}$. But at $\theta_H >10^{\circ}$ the crossover magnetic field is determined by this additional condition and increases due to the factor $\tan\theta_H/\cos\theta_H$.

Interestingly, some results of Figs.~\ref{fig3} and \ref{fig4} can be semi-quantitatively understood if we formally set $\tilde p_0\ll  1$ in the above formulas. In this case  the band-contact helix will look like a practically straight line terminating on the opposite faces of the Brillouin zone, $\lambda$ in Eq.~(\ref{22}) tends to $\cos\theta_H$, $M_{xy}$ defined by Eq.~(\ref{23}) becomes small, and equations (\ref{23}) and (\ref{24}) at $\cot\theta_H \gg \tilde p_0$ transform into
\begin{eqnarray}\label{29}
M_z&\approx& \frac{v_F (2H|\cos\theta_H|)^{1/2}}{\sqrt \pi d}\left(\frac{e}{\hbar c}\right)^{\!3/2}\!\!\!\nu_HK(u), \\
{\bf M}_{xy} &\approx& \frac{\tilde p_0^2}{4}\tan\theta_H {\bf h}_{\perp} M_z, \nonumber
  \end{eqnarray}
where $\nu_H$ is a sign of $\cos\theta_H$, and $u$ is given by Eq.~(\ref{21}) with $\lambda=\cos\theta_H$. At small $u$, one has $K(u)\approx $ const, and $M_z$ is proportional to $|\cos\theta_H|^{1/2}$, whereas $M_{xy}\propto \sin\theta_H/|\cos\theta_H|^{1/2}$; cf. Fig.~\ref{fig3}. Since at $u\gg 1$, $K(u)$ is the oscillating function of $u$,
 \begin{equation}\label{30}
 K(u)\approx 0.5\sqrt{u}(u-[u]-0.5),
 \end{equation}
equations (\ref{29}) show that $M_z$ and $M_{xy}$ oscillate with changing $H|\cos\theta_H|$, cf.  Figs.~\ref{fig4}.

\section{Conclusions}

Whatever the shape of the band-contact line in a topological line-node semimetal, formulas (\ref{5}), (\ref{7}), (\ref{8}), (\ref{10})-(\ref{15}) enable one to calculate the magnetization of this semimetal either as a function of chemical potential or as a function of the charge-carrier density in it. The formulas take into account a dispersion of the degeneracy energy $\varepsilon_d$ along the nodal line, but it is implied in their deriving that the difference $\varepsilon_{max}- \varepsilon_{min} \equiv 2\Delta$ between the maximal and minimal values of $\varepsilon_d$ is essentially less than the characteristic scale  ($\sim 1$ eV) of the electron-band structure of crystals.
In the case of the semimetal with a closed band-contact line lying in a plane perpendicular to an axis of n-fold symmetry, the obtained formulas reduce to those derived in Ref.~\onlinecite{m-sh16}.

At low temperatures the magnetization of the line-node semimetals  generally exhibits the de Haas - van Alphen oscillations, and these  oscillations are shifted in phase as compared to the case of usual metals due to the Berry phase $\pi$ for electron orbit surrounding the band-contact line. This shift is the characteristic property of the topological line-node semimetals. However, the $H$-dependence of the chemical potential can be strong in these semimetals, and the shift of the oscillations can differ from $\pi$, simulating the case of the  Berry phase deviating from this value.

To illustrate the obtained formulas, we apply them to rhombohedral graphite and calculate dependences of the magnetic susceptibility $M_i/H$ of this semimetal on the temperature and the magnetic field,  Figs.~\ref{fig2}--\ref{fig4}. In particular, for magnetic fields  directed along the $z$ axis, we find that the strong dependence of the chemical potential on the magnetic field noticeably changes the shape of the de Haas - van Alphen oscillations and completely masks their phase shift caused by the Berry phase.

\section{Appendix: Electron spectrum in a magnetic field}

Here we justify formulas (\ref{2}) and (\ref{3}).
If $\theta=0$, these formulas exactly describe the electron spectrum in the magnetic field \cite{m-sh}. At a nonzero $\theta$, for points of the line, $p_c$, at which $\varepsilon_d(p_3)$ reaches its minimum $\varepsilon_{min}$ or maximum $\varepsilon_{max}$ values, i.e., where $\varepsilon_d(p_3)$ can be represented in the form $\varepsilon_d(p_3)\approx \varepsilon_{min}+B(p_3-p_c)^2$ or $\varepsilon_d(p_3)\approx \varepsilon_{max}-B(p_3-p_c)^2$ with a positive constant $B$, formulas (\ref{2}), (\ref{3}) approximately hold in the leading order in the small parameter $\eta^2$ \cite{m-sh},
\begin{eqnarray}\label{3a}
\eta^2=\frac{B|\zeta-\varepsilon_c|\tan^2\!\theta}{V^2} \ll 1.
 \end{eqnarray}
Here $V^2\sim (b_{11}b_{22})^{1/2}$, $\zeta$ is the chemical potential of the electrons in a semimetal, $\varepsilon_c=\varepsilon_{min}$ or $\varepsilon_{max}$, and it is implied in Eq.~(\ref{3a}) that $\theta$ is not close to $\pi/2$. Consider now a point of the line, $p_3^{(0)}$, at which $a_3\equiv d\varepsilon_d(p_3)/d p_3\neq 0$. At this point of general position, the spectrum described by Eq.~(\ref{1}) can be formally obtained, setting $b_{33}\to 0$ in the spectrum of the Dirac {\it point}:
\begin{eqnarray}\label{4a}
 \varepsilon_{c,v}\!\!&=&\!\varepsilon_d(p_3^{(0)})\!+\!a_3\delta p_3+{\bf a}_{\perp}{\bf p}_{\perp}\pm E_{c,v},\\
 E_{c,v}^2\!\!&=&\!b_{11}(p_1)^2+b_{22}(p_2)^2+b_{33}(\delta p_3)^2, \nonumber
 \end{eqnarray}
where $\delta p_3\equiv p_3-p_3^{(0)}$.
In the magnetic field ${\bf H}={\bf n}H$ directed along an arbitrary  unit vector ${\bf n}$, the exact spectrum of electrons described by Hamiltonian (\ref{4a}) has the form \cite{m-sh,m-sh16}:
 \begin{eqnarray}\label{5a}
 \varepsilon_{c,v}^l(p_n)\!=\!\varepsilon_d(p_3^{(0)}) +vp_n \pm \left[\frac{e\hbar \alpha_{D} H}{c}\,l+L\cdot (p_n)^2 \right]^{1/2}\!\!\!\!\!\!,
 \end{eqnarray}
where $l=0,1,2,\dots$; $p_n$ is the component of the quasi-momentum along the magnetic field,
\begin{eqnarray}\label{6a}
\alpha_{D}&=&\frac{2R_n^{3/2}}{b_{11}b_{22}b_{33}\tilde{\bf n}^2}, \nonumber \\
L&=&\frac{R_n}{b_{11}b_{22}b_{33}\tilde{\bf n}^4}, \nonumber \\
R_n\!\!&=&b_{11}b_{22}b_{33}\left(\widetilde{\bf n}^2 - [\widetilde{\bf n}\times\widetilde{\bf a}]^2\right), \\
v&=&\frac{(\tilde {\bf a}\tilde{\bf n})}{\tilde{\bf n}^2}, \nonumber
 \end{eqnarray}
and the components of the vectors $\tilde{\bf n}$ and $\tilde {\bf a}$ are defined by the relations:
 \begin{equation}\label{7a}
\tilde{n}_i\equiv \frac{n_i}{\sqrt{b_{ii}}}, \ \ \ \tilde {a}_i\equiv \frac{a_i}{\sqrt{b_{ii}}}.
\end{equation}
In the limit $b_{33}\to 0$, we find
\begin{eqnarray}\label{8a}
R_n\!\!&\approx&b_{11}b_{22}\left (n_3^2(1-\tilde a_{\perp}^2) -a_3^2(\tilde n_1^2+\tilde n_2^2)\right), \nonumber \\
v\!\!&\approx& \frac{a_3}{n_3}, \\
\alpha_D\!\!&\approx&2(b_{11}b_{22})^{1/2}(1-\tilde a_{\perp}^2)^{3/2}n_3 \left (1- \frac{a_3^2 (\tilde n_1^2+\tilde n_2^2)}{n_3^2(1-\tilde a_{\perp}^2)}\right), \nonumber \\
L\!\!&\propto& b_{33} \to 0, \nonumber
\end{eqnarray}
where $n_3=\cos\theta$. Estimating the ratio $a_3^2 (\tilde n_1^2+\tilde n_2^2)/n_3^2(1-\tilde a_{\perp}^2)$, we obtain,
\begin{equation}\label{9a}
\frac{a_3^2 (\tilde n_1^2+\tilde n_2^2)}{n_3^2(1-\tilde a_{\perp}^2)} \sim \frac{a_3^2 \tan^2\theta}{V^2} \sim \frac{\Delta^2\tan^2\theta}{L^2V^2},
\end{equation}
where $2\Delta\equiv \varepsilon_{max}-\varepsilon_{min}$,
and $L$ is the length of the band-contact line in the Brillouin zone. Since $\Delta$ is assumed to be small as compared to the characteristic scale $LV\sim 1$ eV of the electron band structure in the line-node semimetals, the above ratio is small, too, and it does not exceed the parameter $\eta^2$. Hence, $\alpha_D\approx \alpha\cos\theta$ where $\alpha$ is given by Eq.~(\ref{3}). With the relation $vp_n\approx (a_3/\cos\theta)p_n=a_3\delta p_3$, we find that formula (\ref{5a}) reduces to Eq.~(\ref{2}) for a point $p_3^{(0)}$ of general position.

\setcounter{equation}{0}

\vspace{1cm}

\section{Supplemental materials to ``Magnetization of topological line-node semimetals''}

\section{Derivation of formulas for $\Omega$ potential and magnetization}

We start with the common expression for the $\Omega$-potential per the unit volume (see, e.g., \cite{LP,rum_s}):
\begin{align}
  \Omega_H =  - \frac{2T}{\left ( 2\pi\hbar \right )^2} \frac{eH}c
  \sum_{c,v} \sum_{l=0}^\infty{}^{\prime}\int_0^L dp_3\cos\theta \, {\rm ln}
  \left ( 1+ \exp{\frac{\zeta - \varepsilon_{c,v}^l(p_3)}T} \right ) , \label{y0}
\end{align}
where $\zeta$ is the chemical potential, $T$ is the temperature, the electron energy in the magnetic field, $\varepsilon_{c,v}^l(p_3)$, is given by the equations,
 \begin{eqnarray}\label{y1}
\varepsilon_{c,v}^l(p_3)&=&\varepsilon_{d}(p_3) \pm \!\left(\frac{e\hbar\alpha H|\cos\theta|}{c}l\right)^{1/2}\!, \\
\alpha&=&\alpha(p_3)=2(b_{11}b_{22})^{1/2}(1-\tilde a_{\perp}^2)^{3/2}, \label{y2}
 \end{eqnarray}
the prime near the sum means that the term corresponding to $l=0$ is taken with the additional factor $1/2$, the integration is carried out over the length $L$ of the band-contact line in the Brillouin zone, $\theta=\theta(p_3)$ is angle between the magnetic field ${\bf H}$ and the tangent to the band-contact line at the point $p_3$. In Eq.~(\ref{y1}) and formulas below, the signs ``$+$'' and ``$-$'' correspond to the conduction ``$c$'' and valence ``$v$'' bands, respectively. We have also assumed the two-fold degeneracy of these  bands in spin. With Eq.~(\ref{y1}), it is clear that $\Omega_H$ is expressed in terms of the combination  $H_3=H\cos\theta$. We calculate only the part of the $\Omega_H$ that depends on the magnetic field,
\begin{align}\label{y4}
  \Omega(\zeta,H) = \Omega_H - \Omega_0\,,
\end{align}
where $\Omega_0=\lim_{H\rightarrow 0}\Omega_H$ is the $\Omega$-potential at zero magnetic field.
At $T=0$ we obtain from Eq.~(\ref{y0}):
\begin{align}\label{y5}
  \Omega(\zeta,H) =& -\frac{e^{3/2}}{2\pi^2\hbar^{3/2}c^{3/2}}
                     \sum_{c,v} \int_0^L dp_3\sqrt{\alpha(p_3)} H_3\sum_{l=0}^\infty{}^{\prime}
                     \left ( w \pm \sqrt{H_3l} \right )
                     \sigma  \left ( w \pm \sqrt{H_3l} \right ) \nonumber \\
                   & + \frac{e^{3/2}}{2\pi^2\hbar^{3/2}c^{3/2}}
                     \sum_{c,v} \int_0^L dp_3 \sqrt{\alpha(p_3)}\int_0^{\infty} dx \left ( w \pm \sqrt{x} \right )
                     \sigma  \left ( w \pm \sqrt{x} \right ),
\end{align}
where $\sigma(x)=1$ if $x>0$, and $\sigma(x)=0$ if $x<0$, and
\begin{align}\label{y6}
  w \equiv \frac{[\zeta-\varepsilon_d(p_3)] \sqrt{c}}{\sqrt{e\hbar \alpha(p_3)}}.
\end{align}
Formula (\ref{y5}) can be rearranged as follows:
\begin{align}\label{y7}
  \Omega(\zeta, H) = & - \frac{e^{3/2}}{2\pi^2\hbar^{3/2}c^{3/2}} \sum_{c,v} \int_{0}^{L} dp_3 \sqrt{\alpha(p_3)}  H_3  \Bigl [
                        \frac 12 w \sigma(w) - \int_0^{1/2} dl' (w \pm \sqrt{H_3l'})) \sigma (w \pm \sqrt{H_3l'}) \nonumber \\
                      & + \sum_{l=1}^{\infty} \left \{
                        (w \pm \sqrt{H_3l}) \sigma (w \pm \sqrt{H_3l}) -
                        \int_{l-1/2}^{l+1/2} dl' (w \pm \sqrt{H_3l'}) \sigma (w \pm \sqrt{H_3l'}) \right \} \Bigr ]  ,
\end{align}
where we have made the formal substitution $x=H_3l'$.
Using the identity $\sigma(x) + \sigma(-x) \equiv 1$, one can show that
\begin{align}\label{y700}
  \sum_{c,v} (w \pm \sqrt{H_3l}) \sigma (w \pm \sqrt{H_3l}) = 2w + \sum_{c,v} (- w \pm \sqrt{H_3l}) \sigma (- w \pm \sqrt{H_3l}),\ \ \ w\sigma(w)=w-w\sigma(-w).
\end{align}
Hence, $\Omega(\zeta, H)$ in Eq.~(\ref{y7}) does not depend on a sign of the $w$, and we can replace $w$ by $|w|$ in Eq.~(\ref{y7}).
Using also the fact that $\sigma (|w| + \sqrt{H_3l})= 1$, we arrive at
\begin{align}\label{y701}
  \Omega(\zeta, H) = & - \frac{e^{3/2}}{2\pi^2\hbar^{3/2}c^{3/2}} \int_{0}^{L} dp_3 \sqrt{\alpha(p_3)}  H_3 \Bigl [
                       |w| - \int_0^{1/2} dl' (|w|+ \sqrt{H_3l'})) - \int_0^{1/2} dl' (|w|- \sqrt{H_3l'})) \sigma (|w|- \sqrt{H_3l'}) \nonumber \\
                     & + \sum_{l=1}^{\infty} \left \{
                       (|w|+ \sqrt{H_3l})- \int_{l-1/2}^{l+1/2} dl' (|w|+ \sqrt{H_3l'})  \right \} \nonumber \\
                     & + \sum_{l=1}^{\infty} \left \{
                       (|w|-\sqrt{H_3l}) \sigma (|w|- \sqrt{H_3l}) -
                        \int_{l-1/2}^{l+1/2} dl' (|w|- \sqrt{H_3l'}) \sigma (|w|- \sqrt{H_3l'}) \right \}
                       \Bigr ]
\end{align}
Note that the last sum in Eq.~(\ref{y701}) is, in fact, finite due to the factor $\sigma$. Combining all the integrals containing $\sigma (|w|- \sqrt{H_3l'})$, we obtain
\begin{align}\label{y7011}
  \int_{0}^{\infty} dl' (|w|- \sqrt{H_3l'}) \sigma (|w|- \sqrt{H_3l'}) = \int_{0}^{w^2/H_3} dl' (|w|- \sqrt{H_3l'}) = \frac 13 \frac{|w|^3}{H_3},
\end{align}
and
\begin{align}\label{y702}
  \Omega(\zeta, H) = & - \frac{e^{3/2}}{2\pi^2\hbar^{3/2}c^{3/2}} \int_{0}^{L} dp_3 \sqrt{\alpha(p_3)}  H_3 \Bigl [
                       \frac 12 |w| - \sqrt{H_3} \frac 1{3\sqrt{2}}  \nonumber \\
                     & + \sqrt{H_3} \sum_{l=1}^{\infty} \left \{ l^{1/2} -\frac 23 ((l+\frac 12)^{3/2} - (l-\frac 12)^{3/2})  \right \} \nonumber \\
                     & +  \sum_{l=1}^{[u]} \left \{ (|w|-\sqrt{H_3l})  \right \} - \frac{|w|^3}{3H_3}  \Bigr ],
\end{align}
where $[u]$ means the integer part of $u$,
\begin{equation}\label{y8}
u(p_3) \equiv \frac{w^2}{H_3}=\frac{[\zeta-\varepsilon_d(p_3)]^2 c}{e\hbar  \alpha(p_3) H|\cos\theta|}=\frac{cS(p_3)}{2\pi e\hbar H},
\end{equation}
and $S(p_3)$ is the area of the cross section of the Fermi surface by the plane perpendicular to the magnetic field and passing through the point with the coordinate $p_3$.
Thus, equation (\ref{y702}) reduces to the formula:
\begin{align}\label{y71}
  \Omega(\zeta, H) = & - \frac{e^{3/2}}{2\pi^2\hbar^{3/2}c^{3/2}} \int_{0}^{L} dp_3 \sqrt{\alpha(p_3)} \Bigl \{ H_3^{3/2} \Bigl [ \sum_{l=1}^{\infty}
      \left ( l^{1/2}  - \frac 23 \left [ \left (l+\frac 12 \right )^{3/2} - \left (l-\frac 12 \right )^{3/2} \right ] \right )
      - \frac 23 \left ( \frac 12 \right )^{3/2} \Bigr ] \nonumber \\
                     & +H_3|w| \left ( [u] + \frac 12 \right ) - H_3^{3/2} \sum_{l=1}^{[u]} l^{1/2} - \frac 13 |w|^3 \Bigr \}.
\end{align}
Using the relation
\begin{equation}\label{y10}
\zeta(-\frac 12, l) - l^{1/2} = \zeta(-\frac 12, l+1)
\end{equation}
for the Hurwitz zeta function $\zeta(-1/2,x)$,
and the asymptotic expansion of this function at $x\gg 1$ \cite{batem}:
 \begin{eqnarray}\label{b8}
  \zeta(-1/2,x)\!= \!-\frac{2}{3}x^{3/2}\!\!+\!\frac{1}{2}x^{1/2} -\frac{1}{24x^{1/2}} +O(\frac{1}{x^{3/2}}),
 \end{eqnarray}
one can calculate the sums in Eq.~(\ref{y71}),
\begin{align}\label{y9}
   \sum_{l=1}^{\infty}
  \left ( l^{1/2}  - \frac 23 \left [ \left (l+\frac 12 \right )^{3/2} - \left (l-\frac 12 \right )^{3/2} \right ] \right )
  - \frac 23 \left ( \frac 12 \right )^{3/2} - \sum_{l=1}^{[u]} l^{1/2} & = \nonumber \\
  \lim_{M\rightarrow \infty} \Bigl [ \sum_{l=1}^{M}
  \left ( l^{1/2}  - \frac 23 \left [ \left (l+\frac 12 \right )^{3/2} - \left (l-\frac 12 \right )^{3/2} \right ] \right )
  - \frac 23 \left ( \frac 12 \right )^{3/2}  - \sum_{l=1}^{[u]} l^{1/2} \Bigr ] & = \nonumber \\
  \lim_{M\rightarrow \infty} \Bigl [ \sum_{l=[u]+1}^{M} l^{1/2} -\frac 23 (M+\frac 12)^{3/2} \Bigr ] & = \nonumber \\
  \lim_{M\rightarrow \infty} \Bigl [ \sum_{l=[u]+1}^{M} [\zeta(-\frac 12, l) - \zeta(-\frac 12, l+1) ]  -\frac 23 (M+\frac 12)^{3/2} \Bigr ] & = \nonumber \\
  \lim_{M\rightarrow \infty} \Bigl [\zeta(-\frac 12, [u]+1) - \zeta(-\frac 12, M+1)  -\frac 23 (M+\frac 12)^{3/2} \Bigr ] & = \nonumber \\
   \lim_{M\rightarrow \infty} \Bigl [\zeta(-\frac 12, [u]+1)+  \frac 23 (M+1)^{3/2}-\frac 12 (M+1)^{1/2} -\frac 23 (M+\frac 12)^{3/2} \Bigr ] & =
  \zeta(-\frac 12, [u]+1).
\end{align}
Eventually, we obtain the following expressions for the $\Omega$ potential and the magnetization at $T=0$:
\begin{eqnarray}\label{4y}
\Omega(\zeta,H)\!=\!-\frac{e^{3/2}H^{3/2}} {2\pi^2\hbar^{3/2}c^{3/2}}\!\!\int_{0}^{L}\!\!\!\!\!dp_3 |\cos\theta|^{3/2}\sqrt{\alpha(p_3)}K_1(u), \\
  {\bf M}(\zeta,H)\!=\!\frac{e^{3/2}H^{1/2}} {2\pi^2\hbar^{3/2}c^{3/2}}\!\!\int_{0}^{L}\!\!\!\!\!dp_3 |\cos\theta|^{1/2}\nu\!\sqrt{\alpha(p_3)} K(u){\bf t},\label{5y}
  \end{eqnarray}
where
${\bf t}={\bf t}(p_3)$ is the unit vector along the tangent to the band-contact line at a point $p_3$; $\nu=\nu(p_3)$ is a sign of $\cos\theta$;
 \begin{eqnarray}\label{6y}
 K_1(u)\!\!\!&=&\!\!\! \zeta(-\frac{1}{2},\![u]\!+\!1\!)+\sqrt{u}([u]+\frac{1}{2})- \frac{1}{3}u^{3/2}, \\
 K(u)\!\!\!&=&\!\!\! \frac{3}{2}\zeta(-\frac{1}{2},\![u]\!+\!1\!)+\sqrt{u}([u]+\frac{1}{2}), \label{7y}
  \end{eqnarray}

For nonzero temperatures, the $\Omega$ potential and the magnetization $M_i(\zeta,H,T)$ can be calculated with the relationships \cite{rum_s}:
 \begin{eqnarray}\label{9y}
 \Omega(\zeta,H,T)=-\int_{\!-\infty}^{\infty}\!\! d\varepsilon \Omega(\varepsilon,H,0)f'(\varepsilon), \\
 M_{i}(\zeta,H,T)=-\int_{\!-\infty}^{\infty}\!\! d\varepsilon M_{i}(\varepsilon,H,0)f'(\varepsilon),\label{10y}
 \end{eqnarray}
where $f'(\varepsilon)$ is the derivative of the Fermi function,
\begin{equation}\label{11y}
 f'(\varepsilon)=-\left[4T\cosh^2\left(\frac{\varepsilon- \zeta}{2T}\right)\right]^{-1}.
 \end{equation}

\section{Weak magnetic fields}

Consider the expression for the $\Omega$ potential in the limiting case of the weak magnetic field, $H\ll H_T$. Specifically, we shall assume that
 \begin{equation}\label{b0}
 u_T=\frac{T^2 c}{e\hbar  \alpha(p_3) H|\cos\theta|} \gg 1.
  \end{equation}
Inserting formula  (\ref{4y}) into Eq.~(\ref{9y}), interchanging the order of the integrations, and replacing $\varepsilon$ by the variable $\sqrt{u}$ defined by the formula,
\[
 \varepsilon=\varepsilon_d(p_3)\pm \frac{(e\hbar  \alpha(p_3) H|\cos\theta|)^{1/2}}{c^{1/2}}\sqrt{u},
\]
 we arrive at
\begin{eqnarray}\label{b1}
\Omega(\zeta,H,T)\!=\!\frac{e^{2}H^{2}}{\pi^2\hbar c^{2}}\!\!\int_{0}^{L}\!\!\!\!\!dp_3\cos^2\!\theta\alpha(p_3) f'(\varepsilon_d(p_3)) I,
  \end{eqnarray}
where
\begin{equation}\label{b2}
I=\int_0^{\infty}\!\!\!\!\!\!d(\sqrt{u}) K_1(u).
 \end{equation}
In deriving (\ref{b1}), we have replaced $\varepsilon$ by $\varepsilon_d(p_3)$ in the argument of the function $f'(\varepsilon)$. This replacement is based on the assumption that one can choose a constant $u_0$ so that $1\ll u_0 \ll u_T$ (and hence $|\varepsilon(u_0)-\varepsilon_d(p_3)| \ll T$) and at the same time $|I-I(u_0)|\ll |I|$, where
\begin{equation}\label{b3}
I(u_0)\equiv \int_0^{\sqrt{u_0}}\!\!\!\!\!\!d(\sqrt{u}) K_1(u).
 \end{equation}
To justify this assumption and to find $I$, let us calculate $I(u_0)$ which can be rewritten as follows:
\begin{eqnarray}\label{b4}
I(u_0)\!\!\!&=&\!\!\!\!\sum_{n=0}^{[u_0]-1}\!\!\!\!\int_{\sqrt n}^{\sqrt{n+1}}\!\!\!\!\!\!d(\sqrt{u}) K_1(u)\!+ \!\int_{\sqrt{[u_0]}}^{\sqrt{u_0}}\!\!\!d(\sqrt{u}) K_1(u),~~
\end{eqnarray}
where $[u_0]$ is the integer part of $u_0$.
With Eq.~(\ref{6y}), we have
\begin{eqnarray}\label{b5}
&~&\int_{\sqrt{n}}^{\sqrt{n+1}}\!\!\!\!\!\!d(\sqrt{u}) K_1(u)= \zeta(-\frac{1}{2},\!n\!+\!1\!)(\sqrt{n+1} -\sqrt{n}) \nonumber \\ &+&(n+\frac{1}{2})\frac{1}{2}-\frac{1}{12}[(n+1)^2-n^2]=\frac{2n+1}{6} \nonumber \\
&+&\zeta(-\frac{1}{2},\!n\!+\!1\!)(\sqrt{n+1} -\sqrt{n}).
\end{eqnarray}
Taking into account the relation (\ref{y10}), we obtain
\begin{eqnarray}\label{b6}
&~&\sum_{n=0}^{[u_0]-1}\zeta(-\frac{1}{2},\!n\!+\!1\!)(\sqrt{n+1} -\sqrt{n}) = \sum_{n=1}^{[u_0]}\zeta(-\frac{1}{2},\!n\!)\sqrt{n} \nonumber \\
&-&\sum_{n=1}^{[u_0]-1}\zeta(-\frac{1}{2},\!n\!+\!1\!)\sqrt{n}= \zeta(-\frac{1}{2},[u_0])\sqrt{[u_0]} \nonumber \\
&+& \sum_{n=1}^{[u_0]-1}[\zeta(-\frac{1}{2},\!n\!)- \zeta(-\frac{1}{2},\!n\!+\!1\!)]\sqrt{n}= \zeta(-\frac{1}{2},[u_0])\sqrt{[u_0]} \nonumber \\
&+& \frac{[u_0]([u_0]-1)}{2},
  \end{eqnarray}
and
\begin{eqnarray}\label{b7}
\sum_{n=0}^{[u_0]-1}\!\!\!\!\int_{\sqrt n}^{\sqrt{n+1}}\!\!\!\!\!\!d(\sqrt{u}) K_1(u)\!&=&
\zeta(-\frac{1}{2},[u_0])\sqrt{[u_0]} \nonumber \\
&+& \frac{2[u_0]^2}{3}-\frac{[u_0]}{2}.
\end{eqnarray}
Using the asymptotic expansion (\ref{b8}) for $\zeta(-1/2,x)$ at  $x\gg 1$, one can estimate the sum (\ref{b7}) and the last term in the right hand side of Eq.~(\ref{b4}),
\begin{eqnarray}\label{b9}
\sum_{n=0}^{[u_0]-1}\!\!\!\!\int_{\sqrt n}^{\sqrt{n+1}}\!\!\!\!\!\!d(\sqrt{u}) K_1(u)\!&=&-\frac{1}{24} +O(\frac{1}{[u_0]}), \\
\int_{\sqrt{[u_0]}}^{\sqrt{u_0}}\!\!\!d(\sqrt{u}) K_1(u)&\approx& -\frac{\{u\}}{48[u_0]}(1-3\{u\}+2\{u\}^2)=O(\frac{1}{[u_0]}), \nonumber
    \end{eqnarray}
where $\{u\}\equiv u_0-[u_0]<1$. Inserting formulas (\ref{b9}) into Eq.~(\ref{b4}), we eventually find that
\[
 I(u_0)=-\frac{1}{24} +O(\frac{1}{[u_0]}),
\]
$I=-1/24$, and hence
\begin{eqnarray}\label{b10}
\Omega(\zeta,H,T)\!=\!-\frac{e^{2}H^{2}}{24\pi^2\hbar c^{2}}\!\!\int_{0}^{L}\!\!\!\!\!dp_3\cos^2\!\theta\alpha(p_3) f'(\varepsilon_d(p_3)),
\end{eqnarray}
With this $\Omega$ potential, we arrive at linear dependence of the magnetization ${\bf M}=-\partial \Omega/\partial {\bf H}$ on the magnetic field,
\begin{eqnarray}\label{b11}
  {\bf M}(\zeta,H,T)\!=\!\frac{e^{2}H} {12\pi^2\hbar c^{2}}\!\!\int_{0}^{L}\!\!\!\!\!dp_3 (\cos\theta)\alpha(p_3)f'(\varepsilon_d){\bf t}.
  \end{eqnarray}

Finally, it is necessary to emphasize that if we started with Eq.~(\ref{5y}) for the magnetization rather than with Eq.~(\ref{4y}) for the $\Omega$ potential and used the same approach in analyzing the case of weak magnetic fields, we would not obtain the correct expression (\ref{b11}) for the magnetization. This is due to the fact that the integral $\int_0^{\infty}K(u)d\sqrt{u}$ does not converge (the integral  $\int_0^{u_0}K(u)d\sqrt{u}$ oscillates with changing $u_0$, and the amplitude of these oscillations does not tend to zero at large $u_0$).

\section{The de Haas - van Alphen oscillations}

The quantity $u$ defined by Eq.~(\ref{y8}) changes along the nodal line from its minimal value $u_{min}$ to its maximal value $u_{max}$. These extremal values of $u$ correspond to minimal and maximal areas (in $p_3$) of Fermi-surface cross sections by planes perpendicular to the magnetic field. Consider Eq.~(\ref{5y}) in the case when \begin{eqnarray}\label{c0}
u_{min},\ u_{max},\ u_{max}-u_{min}\gg 1.
  \end{eqnarray}

Using Eqs.~(\ref{b8}) and (\ref{7y}), we obtain for large $u$:
\begin{eqnarray}\label{c1}
   K(u)\approx \frac{1}{2}\sqrt{u}(\{ u\}-\frac{1}{2})=- \frac{\sqrt{u}}{2\pi}\sum_{n=1}^{\infty}\frac{\sin(2\pi nu)}{n},
  \end{eqnarray}
where $\{ u\}=u-[u]$, and $[u]$ is the integer part of $u$. Thus, the integrand in Eq.~(\ref{5y}) highly oscillates about zero, and only   the band-contact-line portions located near the points at which $u$ reaches the extremal values give contributions to Eq.~(\ref{5y}).  Let $u$ reach the extremal value $u_{ex}$ at the point $p_3^{ex}$, and let us calculate the appropriate contribution to the magnetization. Near $p_3^{ex}$  we can write the following  expansion for $u(p_3)$:
\begin{eqnarray}\label{c2}
   u(p_3)\approx u_{ex}\pm \frac{1}{2}\left|\frac{\partial^2 S}{\partial p_3^2}\right|\frac{c}{2\pi e\hbar H}(p_3-p_3^{ex})^2 \equiv u_c\pm B (\delta p_3)^2,
  \end{eqnarray}
where $\delta p_3\equiv p_3-p_3^{ex}$, $u_{ex}=u_{min}$ or $u_{max}$, the upper sign corresponds to $u_{min}$ and the lower sign to $u_{max}$. Inserting Eqs.~(\ref{c1}) and (\ref{c2}) into formula (\ref{5y}), we arrive at
\begin{eqnarray}\label{c3}
  {\bf M}(\zeta,H)\!\approx\!-\frac{e} {4\pi^3\hbar^2 c}\!\!\sum_{n=1}^{\infty}\!\frac{|\zeta-\varepsilon_d(p_3^{ex})|(\nu {\bf t})_{p_3=p_3^{ex}}}{n}\!\!\int\!\!\!dp_3\!\left(\sin(2\pi nu_{ex}) \cos[2\pi nB(\delta p_3)^2]\!\pm\!\cos(2\pi nu_{ex})\sin[2\pi nB(\delta p_3)^2] \right )\!.~
  \end{eqnarray}
One may set the infinite limits in this integral over $p_3$. Then, we find
\begin{eqnarray}\label{c4}
{\bf M}(\zeta,H)\!\approx -\left(\frac{e}{\hbar c}\right)^{3/2}\!\!\!\frac{H^{1/2}}{2\sqrt{2}\pi^{5/2}}\left|\frac{\partial^2 S}{\partial p_3^2}\right|^{-1/2}\!\!\!|\zeta-\varepsilon_d(p_3^{ex})|(\nu {\bf t})_{p_3=p_3^{ex}}\sum_{n=1}^{\infty}\!\frac{1}{n^{3/2}}\sin\left(n \frac{cS_{ex}}{e\hbar H} \pm \frac{\pi}{4}\right),
  \end{eqnarray}
where the expressions for $B$, Eq.~(\ref{c2}), and for $u_{ex}$, Eq.~(\ref{y8}), have been inserted; $S_{ex}$ is the area of the  extremal cross section perpendicular to the magnetic field.
Formula (\ref{c4}) describes the de Haas - van Alphen oscillations. In particular, we obtain the following expression for the magnetization component $M_{\parallel}$ parallel to the magnetic field:
\begin{eqnarray}\label{c5}
 M_{\parallel}(\zeta,H)\!\approx -\left(\frac{e}{\hbar c}\right)^{3/2}\!\!\!\frac{H^{1/2}}{2\sqrt{2}\pi^{5/2}}\left|\frac{\partial^2 S}{\partial p_3^2}\right|^{-1/2}\!\!\!\!\!
|\zeta-\varepsilon_d(p_3^{ex})|\,|\cos[\theta(p_3^{ex})]|\! \sum_{n=1}^{\infty}\!\frac{1}{n^{3/2}}\sin\left(n \frac{cS_{ex}}{e\hbar H} \pm \frac{\pi}{4}\right).
  \end{eqnarray}

Compare Eq.~(\ref{c5}) with the well-known formula describing the de Haas - van Alphen effect  at $T=0$ \cite{LP,Sh_s,K,graphite_s},
\begin{eqnarray}\label{c6}
 M_{\parallel}(\zeta,H)\!\approx -\left(\frac{e}{\hbar c}\right)^{3/2}\!\!\!\frac{H^{1/2}S_{ex}}{2\sqrt{2}\pi^{7/2}|m_*|} \left|\frac{\partial^2 S}{\partial p_z^2}\right|^{-1/2}\!
 \sum_{n=1}^{\infty}\!\frac{1}{n^{3/2}}\sin\!\!\left(\!2\pi n( \frac{cS_{ex}}{2\pi e\hbar H}-\gamma) \pm \frac{\pi}{4}\right),
  \end{eqnarray}
where the component $p_z$ is along the magnetic field, $m_*$ is the cyclotron mass, the constant $\gamma$ appears in the semiclassical quantization rule,
 \begin{equation}\label{c7}
 S(\varepsilon)=\frac{2\pi e \hbar H}{c}(n+\gamma ),
 \end{equation}
and is expressed in term of the Berry phase $\Phi_B$ for the appropriate electron orbit \cite{prl_s}:
 \begin{equation}\label{c8}
 \gamma=\frac{1}{2}-\frac{\Phi_B}{2\pi}.
 \end{equation}
If the electron orbit surrounds a band-contact line, $\Phi_B=\pi$ and $\gamma=0$; otherwise $\Phi_B=0$ and $\gamma=1/2$ \cite{prl}.
In Eq.~(\ref{c6}), as in Eqs.~(\ref{5a}) and (\ref{c5}), we completely neglect the electron spin.
In the case a line-node semimetal, one has
\begin{eqnarray}\label{c9}
\gamma=0,\ \ \ \ S_{ex}=2\pi\frac{[\zeta-\varepsilon_d(p_3^{ex})]^2}{\alpha |\cos[\theta(p_3^{ex})]|}, \ \ \ \ m_*=\frac{2[\zeta-\varepsilon_d(p_3^{ex})]}{\alpha| \cos[\theta(p_3^{ex})]|}, \ \ \ \ \frac{\partial^2 S}{\partial p_z^2}=\frac{1}{(\cos[\theta(p_3^{ex})])^2}\frac{\partial^2 S}{\partial p_3^2}.
  \end{eqnarray}
Inserting these expressions into Eq.~(\ref{c6}), we arrive at formula (\ref{c5}).


\begin{thebibliography}{}
\bibitem{comm1} If $\tilde a_{\perp}^2>1$ for a part of the band-contact line, this part should be disregarded in all   subsequent formulas.
\bibitem{comm} In formula (10) of Ref.~\cite{m-sh16}, the numerical coefficient should be $0.11\times 10^{12}$ rather than
$0.7\times 10^{12}$. This correction leads to $H_T\sim 2-200$ Oe at $T=4$ K instead of the estimate $H_T\sim 10-1000$ Oe given in Ref.~\cite{m-sh16}.
\end{thebibliography}
\end{document}